\documentstyle[preprint,aps]{revtex}

\begin{document}
\title{Magnetic field effects in ultracold molecular collisions}
\author{Alessandro Volpi and John L. Bohn \cite{byline}}
\address{JILA and Department of Physics, University of Colorado, Boulder, CO}
\date{\today}
\maketitle

\begin{abstract}

We investigate the collisional stability of magnetically trapped
ultracold molecules, taking into account the influence of
magnetic fields.  We compute elastic and spin-state-changing inelastic
rate constants for collisions of the prototype molecule $^{17}$O$_2$
with a $^3$He buffer gas as a function of the magnetic field and
the translational collision energy.  We find that  spin-state-changing
collisions are suppressed by Wigner's threshold laws as long as the
asymptotic Zeeman splitting between incident and final
states does not exceed the height of the centrifugal barrier
in the exit channel. 
In addition, we propose a useful one-parameter
fitting formula that describes the threshold behavior
of the inelastic rates as a function of the field and
collision energy. Results show a semi-quantitative
agreement of this formula with the full quantum calculations,
and suggest useful applications also to different systems.
As an example, we predict the low-energy rate constants 
relevant to evaporative cooling of molecular oxygen. 

\end{abstract}

\pacs{34.20.Cf, 34.50.-s, 05.30.Fk}

\narrowtext

\section{Introduction}

The probable success of experiments aimed at producing 
magnetically trapped ultracold molecular samples depends heavily 
on the effects of collisional processes.  For example, paramagnetic
alkali dimers can be produced via photassociation of ultracold
atoms \cite{Paul}, but the resulting 
molecules, typically in high-lying
vibrational states, are subject to vibrational quenching
collisions \cite{balfordalPRL98,forkhabaldalPRA99}
which can release a
large amount of energy and dramatically affect the efficiency
of the cooling. Alternatively, cold molecules in their 
vibrational ground states can be produced 
either by thermal contact with
a cold helium buffer gas \cite{weidecguifridoynat98} 
or by Stark slowing, for species
that possess an electric dipole moment 
\cite{Meier,Dineen}.  Collisions are
of obvious importance to buffer-gas cooling (BGC), as well as to 
forced evaporative cooling (EC) that will be required to lower
the temperature of these gases further and achieve
for instance Bose-Einstein condensation (BEC).  Both processes require
large elastic collision rates to thermalize the gas.

So far EC has not been realized in practice for molecules, but the
success of the BGC technique for
the production of cold CaH \cite{weidecguifridoynat98}
and PbO \cite{egoweipatfridoyPRA01} molecules suggests
that in the near future it will be possible
to achieve BEC using cold molecules. This would
open the way for a number of new and fascinating
experiments.

In order to be magnetically trapped, atoms or molecules must
be in a weak-field seeking state, i.e. a state whose
energy increases with the strength of the magnetic field.
For each trappable weak-field seeking state
there is in general a lower-energy untrapped strong-field
seeking state, in which the molecules experience a force
away from the center of the trap.  Collisions can drive transitions
between trapped and untrapped states.  These ``bad'' collisions can cause
heating or atom loss.  It is therefore important to assess
the rate constants for the inelastic collisions.

In a series of previous papers the resilience of molecular oxygen
against spin-changing collisions was investigated, in collisions
of O$_2$ molecules both with a helium buffer gas \cite{bohPRA00}, and with
other O$_2$ molecules \cite{avdbohPRA01}.  These studies found that
spin-changing rates due to spin-rotation coupling could be quite large.
However, in the case of the $^{17}$O$_2$ molecule, where
in the limit of zero field the only allowed
exit channels are energetically degenerate with
the incident channels,  spin-flipping 
transitions require boosting the centrifugal angular
momentum from $L$ = 0 to $L$ = 2, meaning that these
processes are
strongly suppressed by the Wigner's threshold laws 
at collision energies smaller than
the height of the exit channel centrifugal barrier.

The results in Refs. \cite{bohPRA00,avdbohPRA01} considered only 
the case of a vanishing
external magnetic field, which is obviously not the case in experiments
that trap molecules using spatially inhomogeneous magnetic fields.
The present paper therefore explores the role that fields play
in determining  spin-changing
collision rates.  As we demonstrate, the presence of a magnetic field 
causes a Zeeman asymptotic splitting between incident and
exit channels, thus lifting the collision energy higher relative to
the centrifugal barrier in the exit channel, 
and removing the Wigner's suppression.

Studies of spin-changing ultracold collisions in the presence of an
external magnetic field have been performed so far only for
atomic species
\cite{tieverstoPRA93,moeverPRA96,boemoeverPRA96}. 
In this paper, we present a detailed dynamical study 
at cold and ultracold temperatures for the
atom-diatom system $^{17}$O$_2$ $-$ 
$^3$He in a field.  The basic model is described
in Sec. II. In Sec. III we calculate elastic and inelastic rate constants
for collisions involving the lowest-lying trappable state of $^{17}$O$_2$
over a wide range of field values (from 0 up to 5000 gauss), then discuss 
the dependence of the rates on collision energy for several representative
values of the field. This system is of direct relevance to
BGC of molecular oxygen. Generally, it allows us to quantify the
removal of the Wigner's law suppression as the field increases in
strength.  On this basis we determine a simple one-parameter fitting
formula that reproduces the trend with field and energy of the loss rates.
In Sec. IV we use this formula to extend previous results on O$_2$-O$_2$
collisions to estimate the influence of the field on EC of this system.

\section{Theory}

As mentioned in the Introduction, we will consider 
in this paper molecules consisting of two
$^{17}$O atoms, whose
nuclear spin $i$ is equal to $5/2$. We
assume that total spin
{\bf I} = {\bf i}$_1$ + {\bf i}$_2$ is conserved in the collision
and polarized to its maximum value {\bf I} = 5,
implying that the even molecular
rotational states $N$ are separated from the odd ones
\cite{Mizushima}.
We limit the discussions
of this paper to the ``even-$N$'' manifold
of molecular states, which is
more appealing for cooling purposes
\cite{bohPRA00,avdbohPRA01} having a spin 1
paramagnetic ground state.  

The electronic spin {\bf S} of the O$_2$ molecule has
magnitude $S$ = 1 in the electronic ground state
$^3\Sigma_g^-$ we are concerned with throughout this paper.
The angular momentum {\bf S} is coupled to the molecular
rotation angular momentum {\bf N} to give {\bf J}, the total
molecular angular momentum, which assumes the values
$N-1$, $N$ and $N+1$ for $N$ $>$ 0 and is 1 for $N$ 
= 0. The Hamiltonian operator ${\bf \hat{H}}_{{\rm O}_2}$
for molecular oxygen in the presence of an external magnetic 
field $B$ can be written as:
\begin{equation}
\label{O2Ham}
{\bf \hat{H}}_{{\rm O}_2} =
B_e {\bf \hat{N}}^2 + {\bf \hat{H}}_{fs} + {\bf \hat{H}}_{B}
\end{equation}
where $B_e$ is the rotational constant. 
The $^{17}$O$_2$ molecule is considered to be a rigid rotor,
with internuclear distance frozen to the equilibrium
value of $r_0$ = 2.282 bohr (the rigid rotor model
has been shown to be very accurate for this system
at the investigated collision energies
\cite{volboh01}). 
The fine-structure Hamiltonian ${\bf \hat{H}}_{fs}$
and the Hamiltonian ${\bf \hat{H}}_{B}$
for the interaction of the molecule with the external
magnetic field follow the treatment
in Ref. \cite{fremildeslurJCP70}, disregarding
the molecular hyperfine interaction.

The fine structure Hamiltonian is given as \cite{fremildeslurJCP70}
\begin{equation}
\label{Hfs}
{\bf \hat{H}}_{fs} = \left(\frac{2}{3}\right)^{1/2}
\lambda T^2 ({\bf \hat{S}},{\bf \hat{S}}) \cdot 
T^2(\vec{\alpha},\vec{\alpha}) + 
\gamma {\bf \hat{N}}\cdot {\bf \hat{S}}
\end{equation}
where $\vec{\alpha}$ is a unit vector parallel
to the molecular axis and $T^2$ is a second-rank
tensor \cite{vanRMP51,tinstrPR55}.
The fine structure parameters $\gamma$ and $\lambda$ 
have been taken from Ref. \cite{Cazzoli}, where they
have been determined by microwave spectroscopy.

The interaction of the field with the electronic
spin can be expressed as
\cite{curMP65,carlevmilACP70}:
\begin{equation}
\label{HH}
{\bf \hat{H}}_{B} = g \cdot \mu_0 {\bf \hat{S}} \cdot 
{\bf \hat{B}}
\end{equation}
where ${\bf \hat{B}}$ indicates the external magnetic
field, $g$ is the g-factor of the electron
and $\mu_0$ is the Bohr magneton.
Following \cite{carlevmilACP70}
we ignore a small interaction between the field
and the rotational angular momentum.

The matrix elements for ${\bf \hat{H}}_{fs}$ 
and ${\bf \hat{H}}_{B}$ have been given
in Ref. \cite{fremildeslurJCP70}, eqs. (A5) and (A6),
for a Hund's case b basis set.
We note here that the molecular rotational quantum number $N$
is no longer strictly a good quantum number for the molecular states,
because different values of $N$ 
are coupled together by the fine-structure
Hamiltonian ${\bf \hat{H}}_{fs}$ and by the interaction
with the external field. 
The molecular total
angular momentum quantum number $J$ is still a good
quantum number with respect to ${\bf \hat{H}}_{fs}$,
but not with respect to the field
interaction ${\bf \hat{H}}_{B}$ term.
However, its projection $M_J$ on the quantization axis
is still conserved.

Consequently, our basis functions
should be labeled as $| n M_J \rangle$, where 
$n$ is a shorthand index denoting the pair of
quantum numbers ($N, J$) in the field dressed
basis \cite{Mizushima}. However, the coupling between
different $N$'s is weak (the fine-structure coupling
is small compared to the rotational separation)
and $N$ can be considered "almost" a good quantum number. 
Similarly, $J$ is also approximately good for
laboratory-strength magnetic fields,
so that we can use without confusion the label 
$| N J  M_J \rangle$.

Magnetic trapping is strongly related to the
behavior of the molecules in a magnetic field.
The low-energy
Zeeman levels of oxygen are displayed in Fig. 1 for
the even-$N$ species.  (Throughout this paper
we report energies in units of Kelvin by dividing by the Boltzmann
constant $k_B$.  These units are related to
wave numbers via 1 K = 0.695 cm$^{-1}$).
In order to be trapped in the usual magnetic traps
a molecule must be in a weak-field-seeking
state, i.e., one whose energy rises with increasing magnetic
field strength.  Thus the state $|N JM_J \rangle$
= $|0 1\;1 \rangle$ 
is the lowest-lying trappable state of the even-$N$ manifold,
and this is the
state on which we focus our attention below.  This state is indicated
by a heavy line in Figure 1. Higher-lying states with $N$ $\ge$ 2
are energetically forbidden at low temperatures.

It is clear from Fig. 1 that for any trapped state there is
an untrapped, strong-field-seeking Zeeman state at a lower
energy. These states are not merely
untrapped but antitrapped, experiencing a force away from
the trapping region. 
In a magnetic trap
collisions with buffer gas atoms,
or more generally with other molecules, will therefore
ultimately deplete the trap of its molecular population.
The time available for cooling processes like BGC or EC,
as well as the lifetime of a molecular Bose-Einstein condensate,
is therefore limited and knowledge of the rate constants
for spin-flipping collisions is essential to predict their
feasibility. 

In Ref. \cite{bohPRA00} the theoretical
framework for  atom - diatom scattering was
derived in the limit of zero external field, 
along the lines of the model originally due to Arthurs
and Dalgarno \cite{Arthurs,Child}, and properly modified
to incorporate the electronic spin of the oxygen molecule.
Here, the formulation of the scattering problem is further extended
to account for the interaction with the external magnetic field.

The full Hamiltonian operator describing the He - O$_2$ collision is
given by
\begin{equation}
\label{totalH}
{\bf \hat{H}} = -{\hbar^2 \over 2 \mu} \left[ {d^2 \over dR^2}
-{ {\bf \hat{L}}^2 \over R^2} \right] + {\bf \hat{H}}_{{\rm O}_2} + V(R,\theta)
\end{equation}
after multiplying the wave function by $R$ in order
to remove first derivatives.
Here $\mu$ is the reduced mass for the He-O$_2$ system,
$R$ is the modulus of the Jacobi vector joining the atom
to the molecule's center-of-mass, ${\bf \hat{L}}^2$ is the
centrifugal angular momentum operator, and ${\bf \hat{H}}_{{\rm O}_2}$
is the molecular oxygen Hamiltonian defined in (\ref{O2Ham}).
The potential term $V$, depending on both the Jacobi
vector $R$ and on the bending angle $\theta$ that the molecule's axis makes
with respect to ${\bf R}$, accounts for the He - O$_2$ interaction.
We use the $ab-initio$ potential energy surface 
(PES) by Cybulski et al. \cite{Cyb}, which
approximates the true well depth to  $\sim$ 20\%. 

The full multichannel calculation requires casting $V(R,\theta)$
in an appropriate angular momentum basis.
Our field dressed basis for close-coupling calculations is then
\begin{equation}
\label{basis}
|{\rm O}_2(^3\Sigma_g^-) \rangle |{\rm He}(^1S) \rangle
| N J M_J L M_L  {\cal M} \rangle
\end{equation}
where the electronic spin quantum number $S$
is not explicitly indicated being always
equal to 1 in the problem treated here.
The quantum number $L$ stands for the partial wave
representing the rotation of the molecule and the He atom
about their common center of mass, $M_L$ is the projection
of ${\bf \hat{L}}$ onto the laboratory axis, and ${\cal M}$
is the laboratory projection of the total angular momentum,
${\cal M} = M_J + M_L$.
At the collision energies of interest we assume that
the oxygen electronic state and
the helium atom state are preserved and therefore we
suppress the first two kets in equation (\ref{basis}) in the following.

We note here that, at variance with the formulation
in the zero field limit, the total angular momentum of the
system ${\bf {\cal J}}$ (equal to {\bf N} + {\bf S} + {\bf L})
is no longer a good quantum number, because different
${\cal J}$'s are coupled together by the interaction
with the external field. 
This means that the dynamical problem is no longer
factorizable for different values of ${\cal J}$,
thus requiring larger numbers of channels to be treated
simultaneously. The problem is still factorizable
for ${\cal M}$, but in general the number of channels 
to be included for each calculation is much larger
than in the previous case. Numerical details of
the calculations will be given in the next section.

The coupled-channel
equations are then propagated using a log-derivative
method \cite{johJCP73} and
solved subject to scattering boundary
conditions to yield a scattering matrix S:
\begin{equation}
\label{Smatrix}
\langle N J M_J L M_L | S({\cal M})
    | N^{\prime} J^{\prime} M_J^{\prime} L^{\prime} M_L^{\prime} \rangle
\end{equation}
As already noted, the projection of the total angular momentum
${\cal M}$ is still a good quantum number, implying that
$M_J$ $+$ $M_L$ $=$ $M_J^{\prime}$ $+$ $M_L^{\prime}$.
Note that in general each of the quantum numbers $N$, $J$, $M_J$, 
$L$, and $M_L$ are subject to change in a collision,
consistent with conserving ${\cal M}$.

Following Ref. \cite{bohPRA00}, the state-to-state
cross sections are given by:
\begin{equation}
\label{crossec}
\sigma_{N J M_J \rightarrow N^{\prime} J^{\prime} M_J^{\prime}} =
{\pi \over k_{N J M_J}^2 }
\sum_{LM_L L^{\prime} M_L^{\prime}}
|\langle N J M_J LM_L | S-I |
 N^{\prime} J^{\prime} M_J^{\prime} L^{\prime}M_L^{\prime}
 \rangle |^2
\end{equation}
and the corresponding rate coefficients
are given by
\begin{equation}
\label{ratek}
K_{N J M_J \rightarrow N^{\prime} J^{\prime} M_J^{\prime}}
= v_{N J M_J}
\sigma_{N J M_J \rightarrow N^{\prime} J^{\prime} M_J^{\prime}}
\end{equation}
where $v_{N J M_J}$ is the relative velocity of the collision
partners before the collision.  For notational convenience we will
in the following refer to collisions that preserve the incident
molecular quantum numbers as ``elastic,'' and those that change the
quantum numbers as ``loss.''

\section{Results}

In this section we consider elastic and state-changing, 
inelastic rate constants
for the incident channel
$| N\;J\;M_J \rangle$ = $|0\;1\;1 \rangle$.
At the investigated collision energies
(from 1 $\mu$K up to 10 K) there are two open inelastic
channels, namely $| N\;J\;M_J \rangle$ = $|0\;1\;0 \rangle$ and
$|0\;1\;-1 \rangle$, both of which are untrapped (see Fig. 1).

\subsection{Magnetic field dependence}

We begin by computing rate constants for the
elastic and spin-flipping
transitions $| 0\;1\;1 \rangle$ $\rightarrow$ 
$| 0\;1\;-1 \rangle$ and 
$| 0\;1\;1 \rangle$ $\rightarrow$ $| 0\;1\;0 \rangle$
in the low-field limit.
Results in this section refer to 1 $\mu$K collision energy and
are converged using
partial waves up to $L$ = 6 and including the rotational
states $N$ = 0, 2, 4 and 6 for the oxygen molecules. 
The maximum value of $R$ to which the coupled-channel equations
are propagated depends on the strength of the field,
ranging from 600 bohr in the case of smallest field 
to 450 bohr for highest values. These parameters assure
rate constants convergent within less then 5 \%,
which is adequate for our purposes.

In principle, the scattering matrix should be determined for each 
possible value of the projection of the total angular
momentum ${\cal M}$. However, we know that s-wave collisions
dominate the incident channel at ultralow collision energies, which
corresponds for our incident channel to the value ${\cal M}=1$.
We have verified that including in the calculation only the ${\cal M}$ = 1
channels changes the results by less than 1 \% 
at $\mu$K energies; in this section
we therefore include only this contribution. The number of channels
to be propagated according to the given convergent
quantum numbers is then only 205.

Fig. 2 shows elastic and inelastic rate constants
at energy $E=1\mu$K and for low values of the field.
The inelastic rate constants are non zero even at zero field,
as shown in Ref. \cite{bohPRA00}. However, in this limit
the final states are degenerate with the initial state, and
inelastic transitions are strongly
suppressed by the presence of a $d$-wave centrifugal barrier
in the exit channel, whose height is about 0.59 K.
This effect is able to suppress
the molecule loss,
at least as long as the collision energy does 
not exceed the barrier height \cite{avdbohPRA01}.

As soon as a field is applied,
the thresholds are no longer degenerate in energy,
so that the energy in the exit channel is not as far
below the centrifugal barrier.
As a consequence, inelastic transitions are no
longer as strongly suppressed, even in the limit of very
low collision energy. Rather, they increase dramatically even
in a weak field, with rates being boosted 
by 5 or 6 orders of magnitude in a 1 gauss field. On the other hand,
elastic scattering is nearly unchanged by the field.

This sudden increase of the inelastic
transition rates can be reproduced semiquantitatively by applying
the distorted wave Born approximation 
(DWBA) \cite{Child}, as has been successfully
done for the magnetic dipolar interaction
of cold alkali atoms \cite{tieverstoPRA93,moeverPRA96}
as well as in a number of problems in
cold collisions \cite{bohjulPRA99,forbaldalhaghelPRA01}.
The first order DWBA is a simple two state perturbation approximation
applicable in cases where inelastic scattering is weak in comparison
with elastic scattering.  The DWBA expression for the off-diagonal
K-matrix elements is:
\begin{equation}
\label{Kmatr}
K_{if} = - \pi \; \int_0^{\infty} f_i(R) V_{if} (R) 
f_f(R) dR
\end{equation}
where $f_i$ and $f_f$ represent the
energy normalized scattering wave function 
of the initial and final states calculated on the diabatic
potential corresponding to the states involved in the
inelastic transition. $V_{if}$ is the corresponding
diabatic coupling term of the Hamiltonian (\ref{totalH}).  
The first-order DWBA result is
shown as a dotted line in Fig.2.

The DWBA  also yields information on the threshold dependence of
the loss rates on energy and field.  To see this, first note that
the spin-changing processes we are considering are strongly dominated at
low collision energy by s-waves in the incident channel, and
by d-waves in the exit channel.  This change of partial wave is
necessary to conserve angular momentum during a collision
that changes the molecule's spin.
For small values of the magnetic field (for which the
Zeeman splitting does not exceed the height of the
exit channel centrifugal barrier) the exit channel is still
in the threshold regime, whereby the wave functions $f_i$ and $f_f$
in (\ref{Kmatr}) can be approximated by the small-argument limit of 
energy-normalized spherical Bessel functions, 
\begin{equation}
\label{bessel}
f_i \propto \sqrt{ k_i} j_{L_i}(k_i R) \propto (k_iR)^{L_i+1/2},
\;\;\;\;\;
f_f \propto \sqrt{ k_f} j_{L_f}(k_f R) \propto (k_f R)^{L_f + 1/2},
\end{equation}
where $k_i$ and $k_f$ are the incident and final wave numbers
and $l_i$ and $l_f$ are the incident and final partial waves.
Assuming a small-$R$ cutoff to insure convergence of the integral
in (\ref{Kmatr}) with respect to $1/R^6$ singularity in the coupling 
potential $V_{if}(R)$, the energy-dependence of the K-matrix element is
\begin{equation}
\label{prop}
K_{if} \propto k_{i}^{L_i + 1/2} k_{f}^{L_f + 1/2} 
\end{equation}
By considering the relationship between $K_{if}$
and the effective rate constant 
$K_{N J M_J \rightarrow N^{\prime} J^{\prime} M_J^{\prime}}$, it is
straightforward to show that the rate constant
behaves approximately as
\begin{equation}
\label{propE}
K_{N J M_J \rightarrow N^{\prime} J^{\prime} M_J^{\prime}} 
\propto E^{L_i} (E + \Delta M_J g \mu_0 B )^{L_f + 1/2}
\end{equation}
where $E$ is the collision energy, and we have taken
into account that the final kinetic energy in the exit channel
is incremented by an amount $\Delta E_B$ 
= $\Delta {M_J}$ $g$ $\mu_0 B$ 
corresponding to the linear Zeeman shift. 
$\Delta {M_J}$
 (= $M_J$ $-$ $M_J^{\prime}$)
stands for the difference between the initial and final
values of $M_J$ in the two channels involved
in the transition. The actual value of $\Delta E_B$ is modified by
quadratic Zeeman shift, but the linear approximation is
adequate for achieving a simple fitting formula. In the
present case these shifts amount to less than  10\% changes in
the approximated rate constants. 
From eq. (\ref{propE}), considering that in our case
$L_i$ = 0 and $L_f$ = 2, a simple expression for the 
rate constants can be derived:
\begin{equation}
\label{rateK}
K_{N J M_J \rightarrow N^{\prime} J^{\prime} M_J^{\prime}} 
=  K_{0} \left( \frac{E + \Delta M_J g \mu_0 B}{E_0} 
\right)^{5/2}
\end{equation}
where $K_{0}$ represents an overall scaling constant
and $E_0$ is conveniently chosen as the height of the centrifugal barrier 
in the exit, d-wave channel.
In the limit of very low collision energy,
the $\Delta M_J g \mu_0 B$ term obviously dominates over $E$, 
leading to a nonzero
rate constant, as is the case for exothermic collisions.

This simple expression allows us to interpret the
threshold behavior of the rates with the field,
explaining the 5/2 exponential dependence on
$B$ found in our calculation and shown explicitly in 
the bilogarithmic plot for the rate constants (Fig. 3). 
Here the dashed lines represent a fit to the rate constants in the limit
of zero magnetic field, yielding coefficients
$K_{0}$ = 2.73 $\times$ 10$^{-14}$ cm$^3$ sec$^{-1}$
and $K_{0}$ = 1.45 $\times$ 10$^{-14}$ cm$^3$ sec$^{-1}$
for the transition to the final states $| 0\;1\;-1 \rangle$
and $| 0\;1\;0 \rangle$, respectively.  Apart from zeros in the
actual rate constants, the overall trend is indeed 
$K_{N J M_J \rightarrow N^{\prime} J^{\prime} M_J^{\prime}} 
\propto B^{5/2}$.  The zeros in the real rates arise from
interferences between the s-and d-wave radial wave functions, as
we have verified qualitatively by the DWBA.  Nevertheless, the simple
one-parameter expression (\ref{rateK}) provides a reasonable upper bound to 
the complete calculation which, it will be recalled, requires
a calculation involving 205 coupled channels.

The simple formula (\ref{rateK}) holds, of course, only when both
incident and final channels are in the threshold regime.  Assuming
low incident energies, this restriction therefore limits the size of
magnetic field for which (\ref{rateK}) applies.  Namely, this expression
is only useful when the Zeeman energy splitting between
incident and final states remains smaller than
the height of the d-wave centrifugal barrier.  For the channels
considered here, these fields are 2430 gauss and 4860 gauss for
the $|01-1 \rangle$ and $|01 0 \rangle$ final states, respectively;
Vertical arrows in Fig. 3 indicate these field values.
Relation (\ref{rateK}) serves as a useful quick fitting formula
for data at smaller fields, 
and allows us to generalize the results obtained
in this paper also to different systems. 
Possible applications are discussed in section IV. 

When stronger values of the magnetic field are 
considered, the DWBA is no longer
able to reproduce the full calculation,
because the coupling is no longer a weak perturbation.
Strong field interaction
mixes up different channels leading to a much more
complicated picture which can not be explained 
in terms of a simple two state model.  However, in this limit the
loss rate constants seem to be unacceptably large anyway, owing to
the effectiveness of spin-rotation coupling in changing spins.
For example, For $B$ $\sim$ 4800 gauss, the inelastic
rate constant for the $| N\;J\;M_J \rangle$
=  $| 0\;1\;-1 \rangle$ exit channel
exceeds that for the elastic one. This indicates 
that the high field limit fine structure changing
collisions represent a potential limitation
for the success of the collisional cooling processes.

\subsection{Collision energy dependence}

In this section, the dependence of elastic and inelastic
rate constants on the collision energy are analyzed
for the zero field limit and for three representative
values of the field, namely 10, 200 and 4500 gauss.

In Fig. 4.a, we report our results for 
collision energies 1$\mu$K - 1 K, for
the four mentioned values of the field.
Elastic scattering, which is largely determined by s-waves in both
incident and final channels, is weakly affected
by the strength of the field. For spin-changing collisions,
however, the energy dependence changes dramatically, in
accordance with (\ref{rateK}), which is shown in the figure using
dashed lines.  Again the trends are well-represented, 
although the formula overestimates the rates due to the
zeros in the real rates described in the previous section.

As a general trend, we observe that
when the field is increased, the low-energy
inelastic rate constants are substantially 
pushed up towards the elastic ones,
but at higher energies the rates
are less sensitive to the field.
This is better illustrated in Fig. 4.b, where
the total loss rate constants (that is
the sum of the inelastic scattering in the
two exothermic channels) for the different
values of the field are plotted together
with the elastic channel results in zero field
(which, as stressed before, are essentially
independent of the $B$ value).

In the range of collision energies from 1$\mu$K to 1 K
and for low magnetic fields
the rates for elastic collisions remain significantly
higher than the rate for the spin-flipping.  In particular,
at buffer-gas-cooling energies of $\sim 1$K and below, the
inelastic rate constants are in the $10^{-14}$ cm$^3$/sec range,
and remain so even in the presence of a field.  This
result verifies  the suitability of the
$^{17}$O$_2$ molecule for BGC.  Given the comparatively small
uncertainties in the PES \cite{Cyb}, the results
shown here are probably fairly realistic for the He-O$_2$
system.

We have continued the analysis in 
the range of collision energies from 1 to 10 K.
We note that for energies larger than the height
of the centrifugal barrier, the approximation of
including only ${\cal M}$ = 1 in the calculation 
is no longer accurate within a few percent,
as was the case at lower energy. We checked 
that in this energy range the full
calculations including all the possible ${\cal M}$
values provides results which 
differ (in the worst case) by about a
factor of 3 for the inelastic channels and by
about a factor of 5 for the elastic one.
The full calculation is computationally very expensive
for $B$ $>$ 0, as opposed to the case of zero field
where a total-${\cal J}$ basis can be adopted.
Results indicate that
inelastic rates for high collision energies are
not much lower than elastic rates, and of course
become higher still at energies where resonances exist.

At collision energies above 1K both Feshbach and shape resonances
appear in the cross sections, as noted in \cite{bohPRA00}.  We find that
these resonances move somewhat as a function of magnetic field.
Nevertheless, they are sufficiently narrow that they are
completely ``washed out'' by thermal averaging in a gas.  We therefore
present these results as a function of temperature rather than energy.
To this end we assume a Maxwellian velocity distribution of
the collision partners characterized by a kinetic 
temperature $T$. The thermally averaged rate constants
are then expressed as:
\begin{equation}
\label{thermalav}
\bar{K}(T) = \Bigl(\frac{8k_{B}T}{\pi \mu}\Bigl)^{1/2} 
\frac{1}{(k_{B}T)^{2}} 
\int_{0}^{\infty}E \sigma(E) e^{-E/k_{B}T} dE
\end{equation}
where $k_B$ is the Boltzmann constant and $\sigma(E)$
stands for cross sections.
To compute this average, the values of the cross sections
for $E$ $>$ 10 K are extrapolated from their values
at 10 K. 

Averaged rate constants are shown in Fig. 5 for the
same set of field values as in Fig. 4.
The condition for magnetic trapping to be
successful is usually expressed as $K_{el}$ $>$ 10
$K_{loss}$ \cite{bohPRA00rc}, so that we can conclude
that for collisions of $^{17}$O$_2$ with $^3$He this
condition is fulfilled at least for temperatures
up to 1 Kelvin and values of the field for which the
asymptotic Zeeman splitting does not exceeds the
height of the exit channel centrifugal barrier.

\section{Applications}

Even in the absence of detailed information on cold collisions of
a particular molecule, the fitting formula (\ref{rateK}) can be used
as an approximate guide to what the rates might be.  For
example, we can inquire about the prospects for evaporatively cooling
$^{17}$O$_2$ molecules once they have been 
successfully cooled in a first stage 
of BGC.  For this system, the d-wave centrifugal 
barrier height is $\sim 13$ mK,
meaning that the threshold law is expected
to hold for field values smaller than $\sim$ 53 gauss
for transitions that produce one or more molecules in the $| 0\;1\;-1 \rangle$
final state. 

In the case of $^{17}$O$_2$ cold collisions we have access to the zero-field
calculations of Ref. \cite{avdbohPRA01}.  We have fit the energy dependence 
of all the inelastic rate constants 
to yield the dashed curve labeled ``$B=0$'' in Figure 6,
which represents the total loss.
The full calculation (solid line) has some additional features due to 
scattering resonances near zero energy, but this will not affect our
conclusions here.  Based on the zero-field fit, we use (\ref{rateK})
to estimate the loss rates in nonzero field.  The general trends are the same,
namely, the rates rise sharply at low energy, but 
are roughly field-independent at larger energies.

For evaporative cooling to be successful requires, roughly speaking, that 
the ratio of elastic to inelastic collision rates, $K_{\rm el}/K_{\rm loss}$,
exceed 100 \cite{moncorsacmyawiePRL93}.  For the estimated results shown 
in Figure 6, this condition holds 
only at energies below $\sim 1$mK for fields as low as 10 gauss, and 
not at all for near-critical fields of $\sim 50$ gauss.  
Thus evaporative cooling
from an initially buffer-gas-cooled sample may prove trickier than previously
expected.  On the other hand, BGC is characterized by a large number of 
molecules cooled in the initial step; it is possible that a certain loss
can be tolerated, and that a final sample at sub-$\mu$K temperatures
will still hold enough molecules to reach critical phase space density
for BEC. Detailed kinetic simulations are required to determine if this is so.
Alternatively, a recent proposal suggests that NH molecules could be cooled
via Stark slowing to temperatures as low as 1 mK \cite{vanjonbetmeiPRA01}.
In this case EC may work quite well.

For many systems of interest to ultracold studies, there does not
exist any information on spin-changing collisions at low enough
temperatures.  In such cases
it may be possible to make order-of-magnitude guesses anyway.  For example,
suppose a rate constant is known for higher temperature collisions.
In the absence of any other information we could simply assert that 
the rate has the same value at the energy $E_0$ corresponding to the
height of the centrifugal barrier.  The fitting formula (\ref{rateK})
then gives the behavior of this rate at lower temperatures.  
For example, the zero-field rate constant in the first panel of 
Fig. 5 has a value of $10^{-11}$ cm$^3$/sec at a temperature 
of 4 K.  Extrapolating this value down to $E_0=0.59$ K yields
a coefficient $K_0 = 7.2 \times 10^{-14}$ cm$^3$/sec
for the total loss, just a factor
of two higher than the fit to the low-energy calculation.

\section{Conclusions}

One of the main aims of this paper is to understand in a broad sense
collision of paramagnetic molecules in the limit of ultralow
temperatures in the presence of a magnetic field.
This is a part of our effort in showing that molecules
with nonzero spin in the lowest energy state 
(such as $^{17}$O$_2$ investigated here) can be
successfully cooled and used for BEC purposes.

Elastic and inelastic scattering in presence of a magnetic field
for the specific system $^3$He - $^{17}$O$_2$ has been characterized
in detail. Our attention has been focused on
the lowest lying trappable state of the molecule.
This information is immediately relevant for BGC of molecular
oxygen, and suggests that the presence of the field does not
particularly hinder the BGC process. This work extends previous
predictions which referred to the oxygen molecule
in zero field \cite{bohPRA00,avdbohPRA01}, and
definitely assess the theoretical possibility for
trapping this species. 

Moreover, we have illustrated the simple underlying physics of
spin-changing rates in general.  For fields low enough that
both the incident and exit channels are in the threshold regime, 
the rate constants vary according to the Wigner-law dependence
($E$ + $\Delta M_J g \mu_0 B$)$^{5/2}$.  
This insight allows us to make estimates
of rate constants beyond the ones calculated in detail.  In
particular, a small field was found to have a dramatic effect on the
evaporative cooling of $^{17}$O$_2$, which must be taken into account
in future experiments aimed at quantum degenerate molecular gases.

\acknowledgements
This work was supported by the National Science Foundation
and by NIST.  We acknowledge useful discussions with A. Avdeenkov
and J. Hutson.

\begin{figure}
\caption{The lowest-energy Zeeman levels of O$_2$ for the
even-$N$ rotational manifold. These levels are usefully
labeled by the approximate rotational ($N$) and total
spin ($J$) quantum numbers, along with the projection $M_J$
of total spin onto the magnetic field. 
The heavy line indicates the lowest-lying
trappable state $| N\;J\;M_J \rangle$ = $| 0\;1\;1 \rangle$.}
\end{figure}

\begin{figure}
\caption{Rate constants (logarithmic scale) for collisions of 
$^3$He and $^{17}$O$_2$ molecules in the $| N\;J\;M_J \rangle$ =
$| 0\;1\;1 \rangle$
initial state as a function of the magnetic field $B$
and collision energy 1$\mu$K, for
low values of the field.
Solid lines are the complete quantum mechanical calculation, while dotted
lines are results of the distorted wave Born approximation (DWBA).
For each curve the final state of the oxygen molecule is indicated.}
\end{figure}

\begin{figure}
\caption{Rate constants for collisions of $^3$He and
$^{17}$O$_2$ molecules in the $| N\;J\;M_J \rangle$ =
$| 0\;1\;1 \rangle$
initial state as a function of the magnetic field $B$
and collision energy 1$\mu$K (bilogarithmic scale).
Solid lines refer to the exact quantum calculation,
while dashed lines refer to the fitting using
eq. (\ref{rateK}).
For each curve the final state is indicated.
Vertical arrows indicate the values of the field for which
the difference of the asymptotic energies $| 0\;1\;1 \rangle$
$-$ $| 0\;1\;-1 \rangle$ (or $-$ $| 0\;1\;0 \rangle$)
exceeds the height of the centrifugal barrier.}
\end{figure}

\begin{figure}
\caption{(a) Collision energy dependence of elastic and
inelastic rate constants in the range 1$\mu$K - 1 K
(bilogarithmic scale). In each panel the corresponding
value of the magnetic field is indicated.
The solid lines refer to the full quantum
calculations
while dashed lines refer to the inelastic rate constants
calculated using relation (\ref{rateK}), with values of $K_0$
as given in the text. (b) Shows the total loss rate constant for
all four field values simultaneously, to facilitate their comparison. In
(b), only the full quantum calculations are shown.}
\end{figure}

\begin{figure}
\caption{Thermally averaged elastic (solid line)
and total loss (dashed line) rates for $^3$He $-$ $^{17}$O$_2$ collisions
as a function of temperature. In each panel,
the corresponding value of the magnetic field is
indicated.}
\end{figure}

\begin{figure}
\caption{Comparison between elastic and total loss rate constants
for collisions of $^{17}$O$_2$ molecules {\it with each other}, for several
representative values of the field.  The zero-field results (solid line)
are reproduced from Ref. [9].  The field-dependent 
estimates (dashed lines) are based on a fit of eq. (\ref{rateK})
to the zero-field rates. Fitted results are shown only for field values
and collision energy ranges where eq. (\ref{rateK}) is expected to hold.}
\end{figure}

\end{document}